# Data-driven Urban Surface Classification Elucidates Global City Heterogeneity

Yiheng Chen[1,2,3], Wai-Chi Cheng[1,2*], Tzung-May Fu[1,2,4*], Wei Tao[1,2], Aoxing Zhang[1,2], Jimmy C. H. Fung[3,5*], Song Liu[1,2], Lei Zhu[1,2], Xin Yang[1,2]

[1] State Key Laboratory of Soil Pollution Control and Safety, Shenzhen Key Laboratory of Precision Measurement and Early Warning Technology for Urban Environmental Health Risks, School of Environmental Science and Engineering, Southern University of Science and Technology, Shenzhen, Guangdong, 518055, China

[2] Guangdong Provincial Field Observation and Research Station for Coastal Atmosphere and Climate of the Greater Bay Area, Southern University of Science and Technology, Shenzhen, Guangdong, 518055, China

[3] Division of Environment and Sustainability, Hong Kong University of Science and Technology, Hong Kong SAR, China

[4] National Center for Applied Mathematics, Shenzhen (NCAMS), Shenzhen, Guangdong, 518055, China

[5] Department of Mathematics, Hong Kong University of Science and Technology, Hong Kong SAR, China

Corresponding Author: *Wai-Chi Cheng

Email: zhengwz@sustech.edu.cn

Corresponding Author: *Tzung-May Fu

Email: fuzm@sustech.edu.cn

Corresponding Author: *Jimmy C. H. Fung

Email: majfung@ust.hk

**Keywords:** urban surface, urban texture, global city heterogeneity, unsupervised clustering, data-driven urban environmental studies.

**Author Contributions:** Conceptualization: T.M.F; Funding acquisition: T.M.F.; Supervision: T.M.F., J.C.H.F.; Methodology: Y.C., W.C.C., T.M.F.; Investigation: Y.C., W.C.C., T.M.F., W.T., A.Z.; Formal analysis: Y.C.; Software: Y.C.; Data curation: Y.C.; Visualization: Y.C.; Writing-original draft: Y.C., W.C.C., T.M.F.; Writing-review & editing: Y.C., W.C.C., T.M.F., J.C.H.F., W.T., A.Z., S.L., L.Z., X.Y.; Project administration: T.M.F.

**Competing Interest Statement:** The authors declare no conflict of interests.

**Abstract**

Accurate urban surface characterization is essential for environmental modeling, risk assessment, and climate adaptation. However, existing classifications of urban surfaces lack the global consistency and physical detail to fully represent present-day urban heterogeneity. To address this need, we developed a globally unified, Data-driven Urban Environmental Zone (DUEZ) framework. By applying unsupervised clustering to high-resolution (500-m) datasets of building morphology, vegetation, and surface imperviousness, we classified global urban surfaces into 27 DUEZs, representing the exposure setting for approximately 85% of the global population. Compared to the Local Climate Zone scheme, DUEZ framework provides a more detailed representation of urban form, capturing the fine-scale mixing of built and vegetated surfaces in modern cities. Further aggregation of DUEZ patterns revealed nine predominant urban textures globally with regional differences and socioeconomic relevance. The DUEZ framework enhances physical representation of complex urban surfaces in numerical models and establishes a consistent, data-driven basis for global urban environmental studies.

**Main Text**

With over 45% of the global population currently resides in urban areas, a percentage projected to exceed 60% by 2050[1], cities are becoming the primary scene for human exposure to environmental and climate risks, such as air pollution, extreme heat, intense precipitation, and other hazardous weather[2,3]. Effective adaptation to and mitigation of these risks rely on accurate modeling of the urban environment, which in turn requires accurate representation of the complex urban surfaces that shape near-surface airflow, surface-atmosphere exchanges, and local hydrology[4,5,6,7]. However, urban surfaces vary widely across the world due to differences in climate, geography, and historical, cultural, economic, and planning contexts[8,9], reflected in the diverse building forms, vegetation patterns, and surface properties in global cities.

An effective way to represent the complexity of urban surfaces is to classify them into characteristic types. The most widely used example of this is the Local Climate Zone (LCZ) framework[10,11], which classifies land surfaces into 17 types, including 10 built types and 7 land cover types. The LCZ built types are primarily defined by building geometry (building heights and spacings) and surface properties (surface perviousness and vegetated fractions). This framework has enabled the global mapping of urban surfaces using remote sensing data, most notably through the World Urban Database and Access Portal Tools (WUDAPT) initiative[12,13]. By assigning standardized aerodynamic and thermodynamic parameters to each LCZ type[14], the framework has been extensively integrated into weather, climate, and environmental models to simulate the influence of urban surfaces on environmental processes in both current and future climates[15,16,17].

Despite its wide adoption, the LCZ framework has notable limitations. First, its original classification was empirically derived from surveys conducted mostly in western cities between 1976 and 2004[9], potentially introducing a geographical bias. Second, the framework is static; parameters such as the 25-m threshold for "high-rise" buildings, defined decades ago, do not reflect contemporary urban development[18]. Most importantly, the representation of urban surface heterogeneity in LCZ is hampered by its simplified categorization. For example, building morphology is coarsely grouped into high-, mid- and low-rise classes, and the mixing of built and vegetated surfaces is largely considered binary. These simplifications limit LCZ's ability to capture the nuanced, blended gradations of modern urban landscapes[13,19]. Although region-specific adaptations[20] and subclassifications of LCZ types[21] have been proposed, such modifications often lack global transferability or depend on locally restricted data, limiting their broader utility. Meanwhile, applications of deep learning in remote sensing data analysis have yielded a wealth of global, high-resolution datasets characterizing surface features[16,22,23,24], providing powerful new inputs for global urban studies[25,26,27]. To date, however, most applications of these datasets have remained within the LCZ framework. Consequently, a globally consistent, publicly accessible, and objectively derived urban classification framework remains an unmet need in environmental research.

To address this critical need, we developed a globally unified, Data-driven Urban Environmental Zone (DUEZ) framework. By applying unsupervised clustering to global high-resolution datasets of building morphology, vegetation, and surface properties, we classified urban surfaces worldwide into 27 distinct DUEZ types (*Methods*) and produced a global DUEZ map at 500-m resolution for the year 2020. To demonstrate the utility of DUEZ for globally consistent urban analysis, we further classified DUEZ maps to identify nine distinct urban textures globally and examined their regional distributions and socioeconomic relevance (*Methods*). Finally, we discussed how the DUEZ framework can be applied to improve the physical representation of urban surfaces in environmental modeling and support comparative urban studies worldwide.

**Characterization of global urban surfaces by the DUEZ framework**

Fig. 1 presents the 27 distinct DUEZ types derived from our global unsupervised classification (*Methods*), alongside their respective aerial view examples. The DUEZ framework characterizes each 500 m × 500 m urban tile by eight key surface parameters most relevant to near-surface airflow and surface-atmosphere exchange[28] (*Methods*): characteristic building height[29] (CBH), average building height (ABH), standard deviation of building height (SDH), building coverage percentage (BCP), characteristic canopy height (CCH), average canopy height (ACH), vegetation coverage percentage (VCP), and impervious surface percentage (ISP). The different DUEZ types span the wide ranges and varied combinations of these surface parameters in global cities (Supplementary Fig. S1), for instance ABH of 6.1–56 m, BCP of 1.1–40%, ACH of 0.12–10.2 m, VCP of 3.7–86%, and ISP of 1.9–78%.

The 27 DUEZs fall into five feature categories (Fig. 1 and Supplementary Fig. S1). Nine building-dominated zones (BH, BMH, BMHP, BM, BMP, BL1, BL2, BL3, BLP) are characterized by substantial building coverage (BCP > 5%), moderate to high percentages of paved area (ISP > 16.7%), and relatively low vegetation coverage (VCP < 16.7%), typically representing downtown cores or densely populated suburbs; each of these zones is further characterized by buildings of distinct or varied heights, and different densities of low-rise buildings. Three pavement-dominated zones (P1, P2, PV) exhibit high impervious surface fractions (ISP > 33.3%) and correspond primarily to industrial areas or large transportation infrastructure such as airports and parking facilities. Five vegetation-dominated zones (V1, V2, V3, V4, V5) feature high vegetation coverage (VCP > 16.7%) and minimal built cover (BCP < 5%) with varying canopy heights and vegetation coverage; these zones often represent large urban parks or the peripheral green spaces of cities. Four sparsely built or bare zones (SB, SBV, S1, S2) feature scattered buildings with little vegetation, commonly found in arid cities, suburbs, excavation sites, or undeveloped plots. Furthermore, the DUEZ framework identifies six mixed built-vegetated zones (XMV1, XMV2, XLV1, XLV2, XLV3, XLVP), characterized by substantial coverage of both buildings (BCP > 5%) and vegetation (VCP > 16.7%). These mixed types are widespread throughout urban and suburban areas, reflecting the prevalent integration of built and natural elements in contemporary cities.

We mapped global urban surfaces (definition extended here to areas with a population density exceeding 100 persons km$^{-2}$) onto the DUEZ classification at 500 m resolution (*Methods*). Fig. 2 shows the resulting global DUEZ map, which covers 3.3% (4,925,305 km$^2$) of the Earth's land surface, corresponding to the urban environments of approximately 85% of the world's population across approximately 20,000 cities. By total area, vegetation-dominated zones are the most widespread (47.0% of all classified surfaces), followed by sparsely-built or bare zones (37.2%); together, building-dominated and mixed zones account for only 13.6% of all classified surfaces. A population-weighted analysis, however, reveals a markedly different exposure profile: building-dominated (16.5%) and mixed (12.8%) DUEZs represent substantially greater fractions of the population-weighted urban area, reflecting that the everyday experiences for urban residents in high population density areas are shaped by building-dominated and mixed landscapes.

Regional DUEZ maps further reveal distinct geographic dependencies in urban surface composition (Fig. 2). Large, contiguous urban expanses are most prevalent in East Asia, the Indo-Gangetic Plain, and parts of Indonesia, whereas urban areas in other regions are more fragmented. The distribution of vegetation-dominated DUEZ types exhibits strong climatic dependence: tropical and subtropical cities mostly feature densely vegetated surfaces (e.g., Java Island, southern China, southeastern Brazil), while temperate cities typically feature more open, lower-canopy vegetation (e.g., Europe and North America). Arid regions (e.g., the Middle East) and areas where urban and agricultural land intermix (e.g., northern China, the Indo-Gangetic Plain) are dominated by sparsely built or bare surfaces.

A key insight from the DUEZ framework is the identification of mixed built-vegetated zones as a distinct feature category. These mixed types are mostly found in suburban and peripheral areas of large city clusters, particularly in East Asia and North America. Their widespread occurrence reflects two concurrent processes: the outward expansion of built-up areas into previously vegetated landscapes[30], and the intentional integration of green infrastructure, such as urban parks and reserved lands, into the urban fabric. Regional differences in the prevalence of mixed DUEZ types further suggest that, in developed regions, urban planning increasingly emphasizes a balance between built-up and green space, while in rapidly developing regions, urban expansion remains largely driven by the conversion of land to built-up uses.

**Enhanced Representation of Urban Heterogeneity by the DUEZ Framework**

By characterizing urban surfaces with greater physical detail than the LCZ framework, the 27-type DUEZ framework more accurately represents modern urban surface features. Supplementary Fig. S2 shows the co-occurrence matrix between DUEZ and LCZ13 classifications for global urban surfaces at 500-m resolution for the year 2020. The analysis reveals that each of LCZ's 10 built types corresponds to between four and eight distinct DUEZ types. For example, areas broadly classified as LCZ 1 (compact high-rise) are further differentiated in DUEZ as zones dominated by various building height ranges (BH, BMH, BM, and BL3) and zones representing distinct mixtures of mid-rise buildings and vegetation (XMV1 and XMV2). Similarly, the LCZ "open built" types (LCZs 4, 5, and 6) are differentiated into DUEZ types that distinguish varying vegetation densities (XLV1, XLV2, V1, V2, and V4), building heights (XMV1, BH, BMH, BM, and BL1), and proportions of bare or paved surfaces (SBV, S1, BMHP, and XLVP).

Beyond built surfaces, the DUEZ framework also provides finer granularity for vegetated urban surfaces, distinguishing five degrees of vegetation density (V1 to V5, and S1) compared to the broader LCZ types A (dense trees) and B (scattered trees). Conversely, several LCZ classes for bare or sparsely vegetated land (LCZs C to F) are consolidated into a more generalized DUEZ category (S1, bare surface). This selective reallocation of detail preferentially enhances the representation of built and vegetated features most relevant to urban microclimate and surface-atmosphere exchange processes.

The enhanced granularity of DUEZ better elucidates the intra-urban heterogeneity in contemporary global cities. A visual comparison of DUEZ and LCZ maps for ten representative global cities and metropolitan regions (Fig. 3) highlights three key improvements. First, DUEZ more precisely delineates the fine-scale juxtaposition of buildings, vegetation, and mixed surfaces within dense urban cores (e.g., Cairo, Riyadh, São Paulo, Moscow, and Buenos Aires). Second, DUEZ better resolves small, fragmented urban patches which are often indistinguishable in coarser classifications (e.g., outskirts of Delhi, Gauteng City-Region, and Riyadh). Third, DUEZ more effectively captures the morphological gradients and functional connections among city centers, suburbs, and satellite towns within greater metropolitan regions (e.g., the Yangtze River Delta, Tokyo metropolitan area, and Greater Los Angeles area).

**Global urban textures as manifested by the DUEZ framework**

Our high-resolution DUEZ maps captured the intra-urban heterogeneity associated with different functional areas within a modern city or city cluster. This spatial composition, referred to in urban planning literature as "urban texture" or "urban fabric", reflects the spatial organization of urban morphology and socioeconomic activities within a city and is crucial for understanding a range of urban environmental or socioeconomic issues, such as transportation and energy use efficiency[31], urban heat environment[32], pollution exposure[33], socio-spatial segregation[34], and environmental justice[35].

We applied unsupervised clustering to the DUEZ maps of 1,600 global cities, revealing nine distinct textures of global cities (Methods): extensive high-rise, extensive low-rise, fragmented high-rise, fragmented low-rise, diffusive, filamentous, nodal, reticular, and irregular. Fig 4 presents the nine urban textures, along with representative DUEZ maps and satellite imagery. For each urban texture, Supplementary Figs. S3 and S4 show the relative frequency of individual DUEZ types and their spatial adjacency patterns, respectively. Extensive cities are characterized by large, contiguous expanses of built and mixed DUEZs, with few bare or sparsely built zones. In contrast, fragmented cities consist of scattered building patches that are loosely connected by mixed surfaces of middle- and low-rise buildings and vegetation. Diffusive cities are predominantly vegetated, interspersed with low-density built areas and bare lands. Filamentous cities feature corridors of low buildings or paved and bare surfaces that often align with major roads or rivers. Both nodal and reticular cities are dominated by bare lands, but nodal cities contain a single, compact urban core, whereas reticular cities comprise multiple small nodes of low-rise buildings connected by networks of low buildings or paved surfaces. Irregular cities exhibit highly variable spatial distributions of buildings, often strongly influenced by local topography such as coastlines or mountains.

Within most types of urban textures, building-dominated and mixed DUEZs are often self-connected, forming cohesive conglomerates of similar functional areas (Supplementary Fig. S4). Two types of urban textures stand out as exceptions. In nodal cities, the built-up areas are not connected, reflecting a lower degree of functional interaction across spaces. Similarly, in extensive low-rise cities, built-up zones are often separated by intervening green spaces, reducing the spatial continuity of buildings. These distinct connectivity patterns suggest that population exposure to urban environments and accessibility to green spaces and other key functional areas vary systematically across different urban textures.

We further observe that the spectra of urban textures vary significantly across regions and with respect to coastal proximity (Fig. 5). Approximately half of inland cities in North America exhibit an extensive low-rise form, a pattern rarely observed elsewhere. Inland cities in Europe and South America show similar profiles, with roughly half characterized as extensive high-rise, followed by diffusive forms. Inland cities in China and the Middle East also share a notable prevalence of filamentous cities; in China these are followed by extensive high-rise cities, whereas in the Middle East reticular and fragmented low-rise cities are more common. Inland cities in India and Sub-Saharan Africa exhibit greater diversity than those in other regions. In India, more than two-thirds of cities are diffusive, fragmented low-rise, and reticular forms. Inland cities in Sub-Saharan Africa take up all textural forms, except extensive low-rise. Indonesia stands out, with over 70% of its inland cities classified as extensive high-rise. Globally, coastal cities tend to be more fragmented or irregular than inland cities within the same region.

Overall, more than half of inland cities in the Global North conform to extensive structures, whether high-rise or low-rise. By contrast, inland cities in the Global South exhibit greater textural diversity, including extensive high-rise, filamentous, reticular, diffusive, and other forms. Notably, extensive low-rise cities are almost exclusively in the Global North, while nodal and reticular cities are largely present in the Global South.

Beyond their environmental implications, the nine distinct urban textures also reflect pronounced socioeconomic disparities, as evidenced by the wide variations in gross domestic product (GDP) at purchasing power parity36 across the 1,600 sampled global cities (Supplementary Figs. S5 and S6; Methods). Extensive low-rise cities, which occur almost exclusively in North America, exhibit the highest mean GDP (40 billion USD). Extensive high-rise and filamentous cities show comparable mean GDP values and similarly wide ranges, though the drivers of this similarity remain unclear given their spatially disparate regional distributions (Fig. 5). Nodal cities display the lowest mean GDP globally (2.9 billion USD), with relatively limited internal disparity due to their scarcity. The widest GDP range is observed among fragmented low-rise cities: the wealthiest city of this type (southwestern Istanbul) has a GDP of 201 billion USD, while the poorest (Saint-Marc, Haiti) has a GDP of less than 10 million. The underlying causes of these wealth disparities across urban textures warrant further investigation. Nevertheless, these patterns illustrate the potential utility of the DUEZ framework and its derived urban textures for broader socioeconomic inquiry.

**Conclusions and implications**

We developed and applied a globally consistent, Data-driven Urban Environmental Zone (DUEZ) framework, providing a high-resolution characterization of urban surface composition and urban morphology worldwide. This framework offers a physically detailed alternative to the widely used LCZ system, capturing the full spectrum of contemporary urban forms, especially the prominence of mixed built-vegetated surfaces, and better delineating the heterogeneity and connectivity of functional areas. The fine-scale morphological information derived from DUEZ enables improved parameterizations for surface energy budgets, near-surface airflow, and surface-atmosphere exchange, which can be integrated into environmental models to enhance simulations of weather and climate across scales.

The DUEZ framework's power can be further enhanced by incorporating temporal and multidimensional data. Applying the DUEZ framework to historical datasets would allow for modeling the temporal evolution of urban forms, providing critical insights for forecasting future development trends. Furthermore, integrating socioeconomic data beyond GDP, such as population health metrics or energy consumption statistics, could offer a more holistic understanding of the complex interactions between urban form and societal function, underscoring DUEZ's value as a foundational dataset for interdisciplinary urban research. DUEZ also serve as a powerful tool for evidence-based urban planning for climate and ecological resilience. Policymakers can leverage DUEZ's high-resolution maps and urban texture information to develop greening initiatives to combat urban heat or design green corridors that connect existing parks, thereby improving biodiversity and ensuring more equitable access to nature for residents.

Finally, unlike static classification schemes, the DUEZ framework is inherently adaptable. Its training procedure is fully reproducible using open-access data, allowing the classification to be updated as urban forms evolve. Users may also customize the framework by incorporating alternative data sources or adjusting feature weights, making DUEZ a flexible and enduring tool for advancing environmental modeling and supporting sustainable urban development in a rapidly urbanizing world.

**Methods**

**Globally unified data-driven classification of urban environmental zones at 500 m resolution**

To classify global urban areas objectively, we first identified candidate regions by selecting 15-arcminute × 15-arcminute land surface blocks with population densities exceeding 500 inhabitants per km2. Using the population data from the publicly available GPW-v4 dataset37, we obtained an initial pool of over 6,000 blocks worldwide. From this pool, a subset of 1,600 blocks was randomly selected for analysis. To ensure geographic representativeness and mitigate regional sampling bias, blocks with insufficient source data were excluded, and selection within any 20° × 20° region was capped at 100 blocks. This sampling strategy ensures the dataset encompasses the global diversity of urban surfaces, including varied building morphology, vegetation patterns, and surface imperviousness.

Previous studies have shown that urban morphological and climate statistics become approximately homogeneous when spatially averaged over a characteristic scale of 400–500 m, effectively filtering out the microscale heterogeneity introduced by individual buildings 9,38. Therefore, we subdivided each sampled 15-arcminute block into 500 m × 500 m tiles, compiling a training dataset of over 2 million individual urban tiles. Each tile was characterized by eight key parameters most relevant to near-surface airflow and surface-atmosphere exchange28 (Supplementary Table S1). Four building metrics, namely characteristic building height29 (CBH), average building height (ABH), standard deviation of building height (SDH), and building coverage percentage (BCP), were derived from the 3D Global Building Footprint (3D-GloBFP) dataset for the year 2020 (native resolution: approximately 10 m; height uncertainty: 1.2 m)24. Three vegetation metrics, namely characteristic canopy height (CCH), average canopy height (ACH), and vegetation coverage percentage (VCP), were calculated from the High-Resolution Canopy Height Model (HR-CHM) dataset for 2018-2020 (native resolution: 0.6 m; height precision: 1 m; absolute error: 2.8 m)39. The impervious surface percentage (ISP) was obtained from the Global Impervious Surface Dataset (GISD30) for 2020 (native resolution: 30 m)40. Within the training data, the three building-height parameters (CBH, ABH, SDH) were correlated ($R^2$ = 0.59 to 0.88; Supplementary Fig. S7), as were the two canopy height metrics (CCH and ACH, $R^2$ = 0.67; Supplementary Fig. S7). Canopy height also exhibited correlation with vegetation coverage (VCP), likely reflecting the characteristic densities of different plant functional types (e.g., trees vs. shrubs) in urban settings. Despite these correlations, we retained all eight parameters to preserve a complete physical description of the urban surface features for classification.

We applied k-means clustering to objectively classify the over 2 million urban surface tiles constituting the training dataset. All input parameters were normalized using min-max scaling. Three normalized variables with more leptokurtic (sharply peaked, heavy-tailed) global distribution (CBH, ABH, SDH) were assigned greater weightings in the clustering algorithm (Supplementary Table S1). We used the Calinski-Harabasz score to evaluate the hyper-parameter k, which was the number of optimal clusters. The score increased continuously for k values up to 50, indicating improved feature separation with greater granularity; however, overly fine subdivisions would reduce categorical utility for urban analysis. To balance, we selected k = 40 and merged clusters with similar centroid values of surface parameters, using mean population density and satellite imagery as auxiliary guidance. This process resulted in the final identification of 27 distinct DUEZ types (Fig. 1).

Finally, we mapped all global urban surfaces (here expanded to all global 15-arcminute land surface blocks with population densities exceeding 100 inhabitants per km2) to the 27 defined DUEZ types at 500-m resolution. For each 500-m tile, we projected its eight surface parameters into the DUEZ feature space and assigned the DUEZ type with the least Euclidean distance. Tiles with insufficient source data for the surface parameters were excluded. The resulting global 500-m DUEZ dataset included 26,617,213 tiles, covering an area exceeding 5 million km2. This dataset represents approximately 3.3% of the global land surface but reflects the urban environmental exposure scenes for approximately 85% of the world's population (Fig. 2).

To compare the characterization of urban surfaces using the DUEZ and LCZ frameworks, we calculated the co-occurrence matrix of global DUEZ and LCZ maps at 500-m resolution. The global LCZ map (100-m native resolution) was from the open-access LCZ Generator13 and resampled to 500 m using a majority filter.

**Classification of global urban textures**

Building on the global DUEZ map, we next classified urban textures worldwide. We represented each of the 1,600 densely populated 15-arcminute land surface blocks as a 25 km × 25 km 2D DUEZ map. Each map was analogous to a city-scale "image", wherein the DUEZ type of a 500-m tile corresponded to the color of a pixel; unclassified tiles were assigned null values. In a few cases, our coordinate-based separation of city blocks would divide an administratively contiguous city into two blocks; we accepted that objective delineation to ensure methodological consistency.

We developed a deep-clustering algorithm to classify these 1,600 city images. Each 2D DUEZ array was encoded into a latent representation using a convolutional neural network-based autoencoder, flattened into a 1D feature vector, and clustered via k-means. We used several metrics to evaluate the optimal k-value. Mathematically, the classification performance generally decreased with increasing k, but too few clusters lacked practical utility for distinguishing urban forms. We examined the clusters for k values of 3 to 12 and determined that 9 clusters gave the best contrast of urban textures and regional differences (Figs. 3 and 4). Satellite images representative of each of the nine urban textures are from the Sentinel-2 MSI Level-1C images, obtained from the Copernicus Data Space Ecosystem (https://dataspace.copernicus.eu/)41, covering the period from 2025-02-28 to 2025-12-17.

**Evaluation of Gross Domestic Product differences among distinct urban textures**

To demonstrate the utility of our DUEZ framework and urban texture classification for socioeconomic inquiry, we analyzed differences in Gross Domestic Product (GDP) across the nine identified global city structural types. We used global gridded GDP data for 2020 at 30-arcsecond resolution36. To control for the influence of varying city sizes, we restricted the analysis to a consistent 25 km × 25 km spatial domain for each city. A Kruskal-Wallis non-parametric test was first applied to determine whether GDP differed significantly among the nine urban textures. Subsequently, Dunn's post-hoc test was applied to make pairwise comparisons.
If your research involved human or animal participants, please identify the institutional review board and/or licensing committee that approved the experiments. Please also include a brief description of your informed consent procure if your experiments involved human participants.

**Acknowledgments**

This work was supported by the National Natural Science Foundation of China (42325504, 42461160326, 42375193, 42305188), the Shenzhen Science and Technology Program (KQTD20210811090048025), and the High-level Special Funds (G03034K006). Computational resources were supported by the Center for Computational Science and Engineering at the Southern University of Science and Technology. We acknowledge the European Space Agency (ESA) and the Copernicus Data Space Ecosystem for the Sentinel-2 MSI data.

**Figures and Tables**

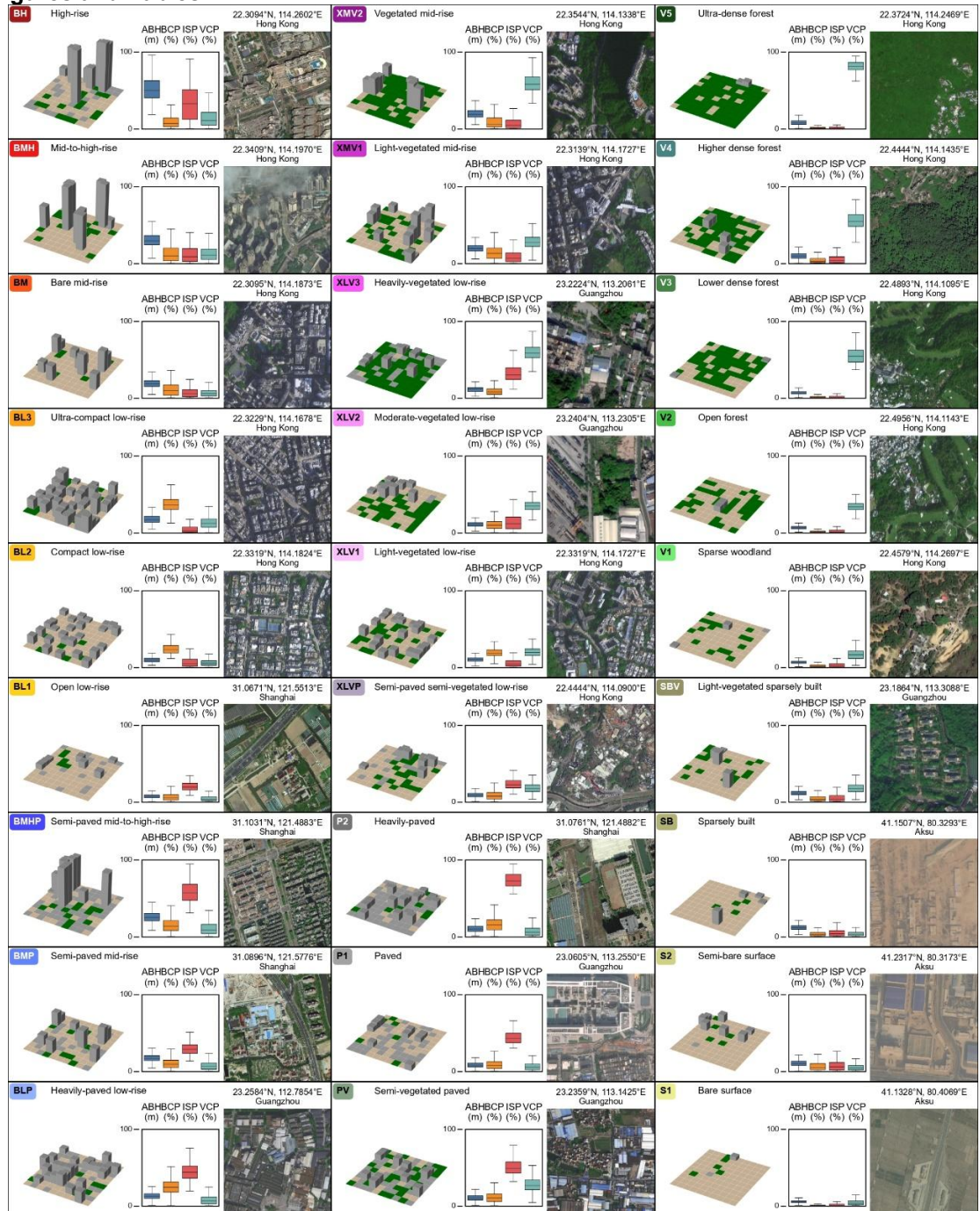

**Figure 1.** 27 DUEZ types and examples of their respective aerial views. Each type is illustrated with a 3D schematic depicting the coverage of buildings and impervious surfaces (gray), vegetated surfaces (green), and bare surfaces (beige). Box plots show the characteristic ranges of four key parameters: average building height (ABH), building coverage percentage (BCP), impervious surface percentage (ISP), and vegetation coverage percentage (VCP). An example aerial image from LocaSpace Viewer (http://www.locaspace.cn/) is provided for each zone, accompanied by the corresponding city name and geographic coordinates.

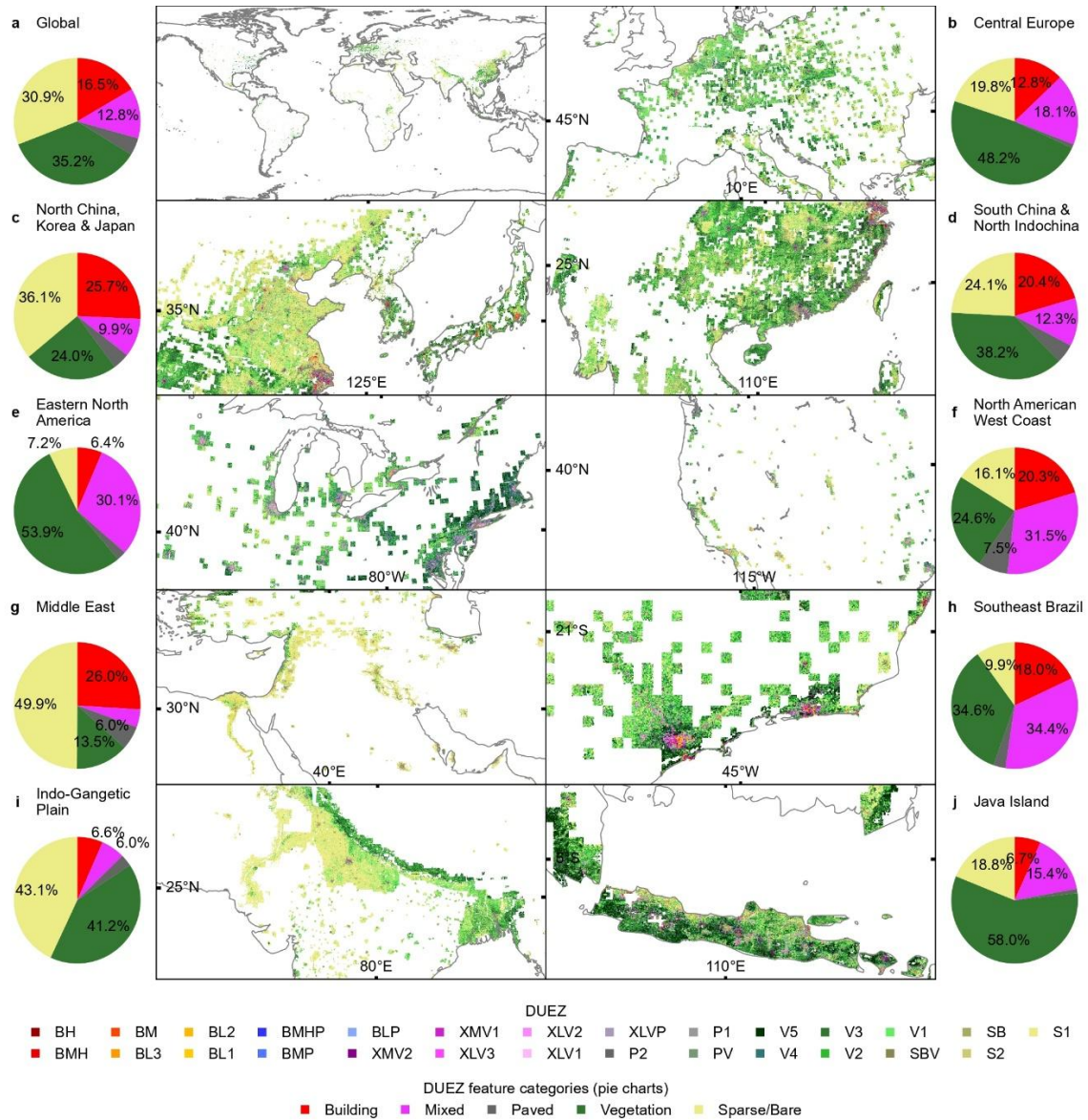

**Figure 2.** Global and regional distributions of DUEZ types and functional categories.
a, DUEZ map for the World. b-j, DUEZ maps for nine representative regions. Each pixel is colored by its DUEZ type and was sampled at 2 km resolution from the underlying 500 m DUEZ dataset; pixels with insufficient source data or with <100 people km-2 are left blank. The accompanying pie charts indicate the percentages of population-weighted classified surface area occupied by each of the five DUEZ feature categories.

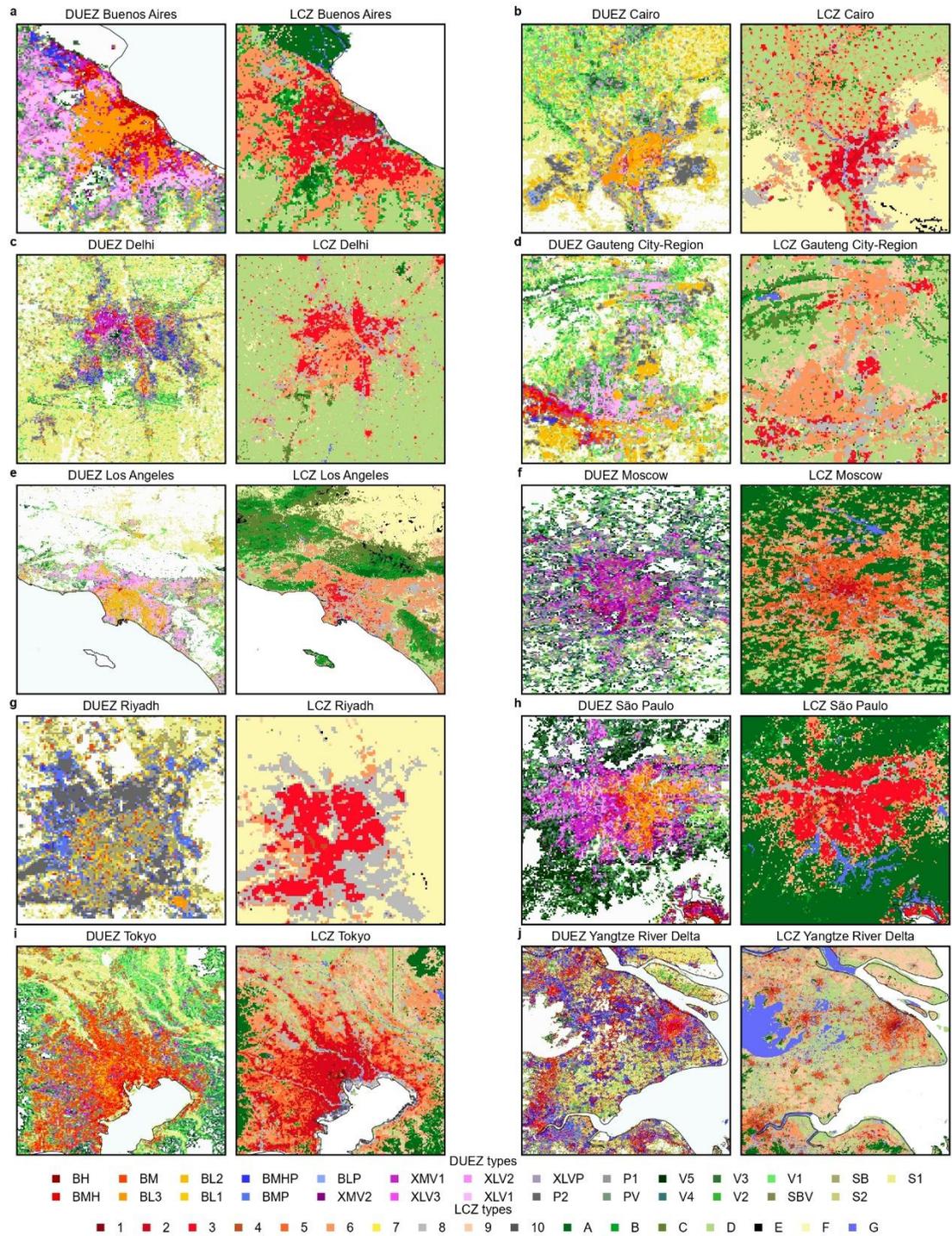

**Figure 3.** Comparison of urban heterogeneity represented by the DUEZ and LCZ frameworks in 10 major cities or city clusters.
a, Buenos Aires, Argentina. b, Cairo, Egypt. c, Delhi, India. d, Gauteng City-Region, South Africa. e, Greater Los Angeles region, U.S. f, Moscow, Russia. g, Riyadh, Saudi Arabia. h, São Paulo, Brazil. i. Tokyo metropolitan area, Japan. j, Yangtze River Delta region, China. For each city, the left and right panels indicate DUEZ and LCZ classifications, respectively.

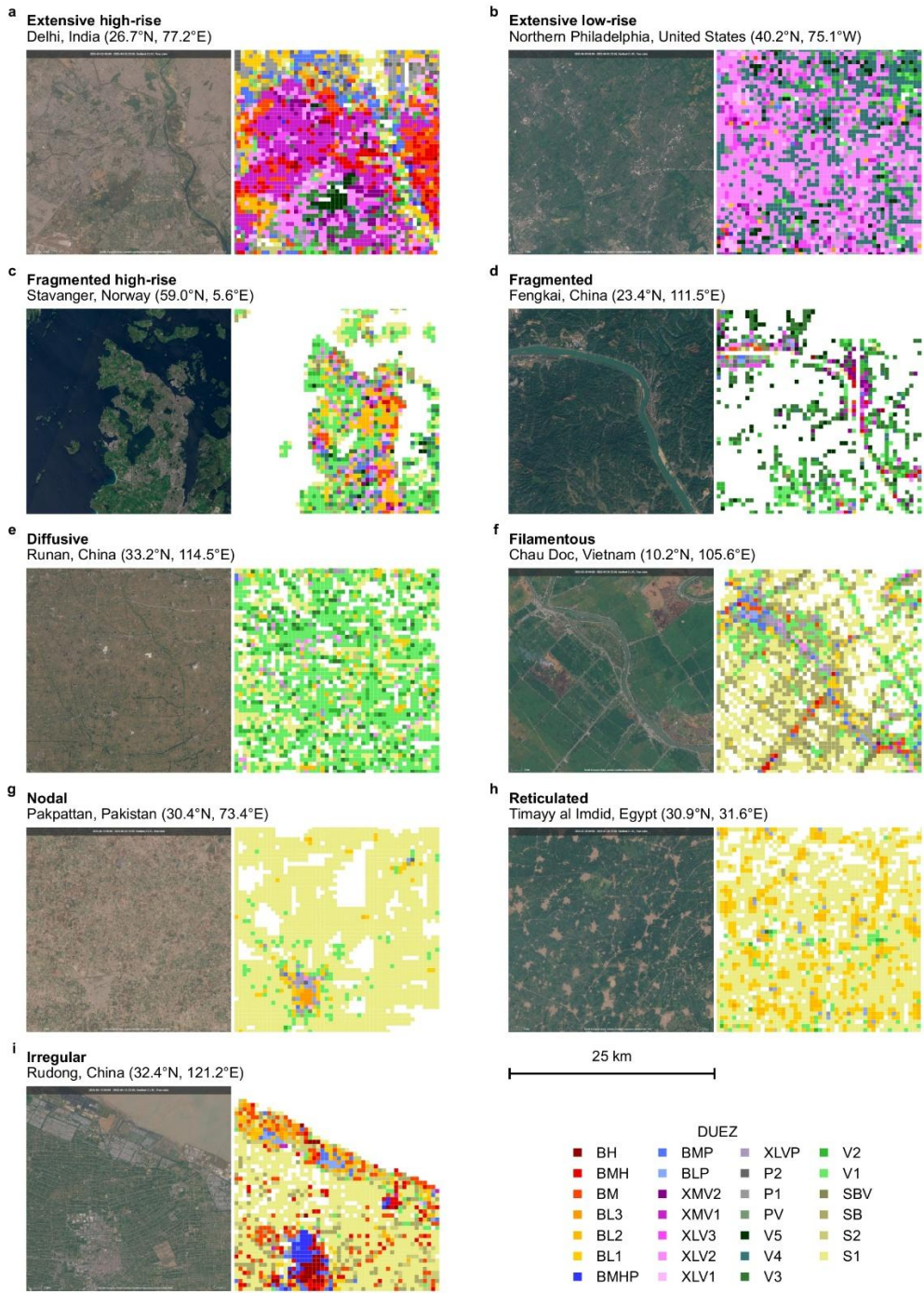

**Figure 4.** Representative satellite images and DUEZ maps of nine distinctive urban textures revealed by the DUEZ framework.
a, extensive high-rise. b, extensive low-rise. c, fragmented high-rise. d, fragmented low-rise. e, diffusive. f, filamentous. g, nodal. h, reticular. i, irregular. Each satellite image and the corresponding DUEZ map cover a 25 km × 25 km area.

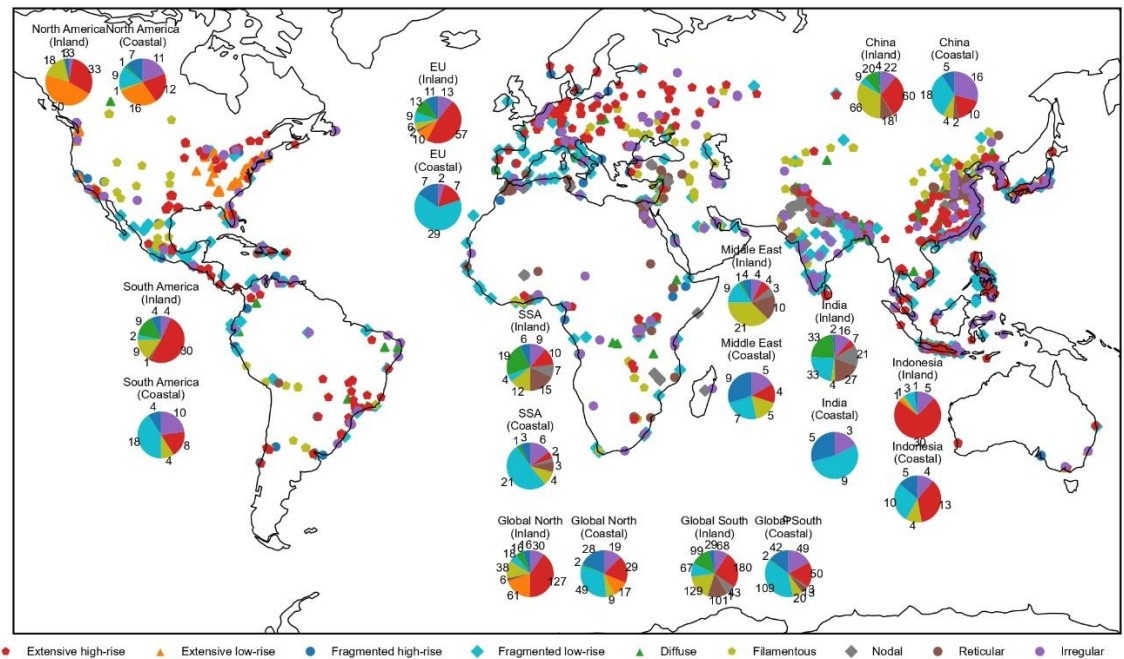

**Figure 5.** Global and regional spectra of urban textures in 1,600 global cities.
Symbols indicate the urban textures of cities as determined by their DUEZ maps. Color-coded pie charts indicate the spectra of urban textures for various regions of the World; numbers indicate city counts.

**Fig. S1: Box plots of characteristic surface parameters distinguishing the 27 DUEZ types.**
**a**, characteristic building height (CBH, in m). **b**, average building height (ABH, in m). **c**, standard deviation of building height (SDH, in m). **d**, building coverage percentage (BCP, in %). **e**, impervious surface percentage (ISP, in %). **f**, characteristic vegetation canopy height (CCH, in m). **g**, mean vegetation canopy height (ACH, in m). **h**, vegetation coverage percentage (VCP, in %). Red and black bars indicate median and mean values, respectively. Black whiskers indicate standard deviations. Blue boxes and whiskers indicate quartiles and inter-quartile ranges (IQR), respectively.

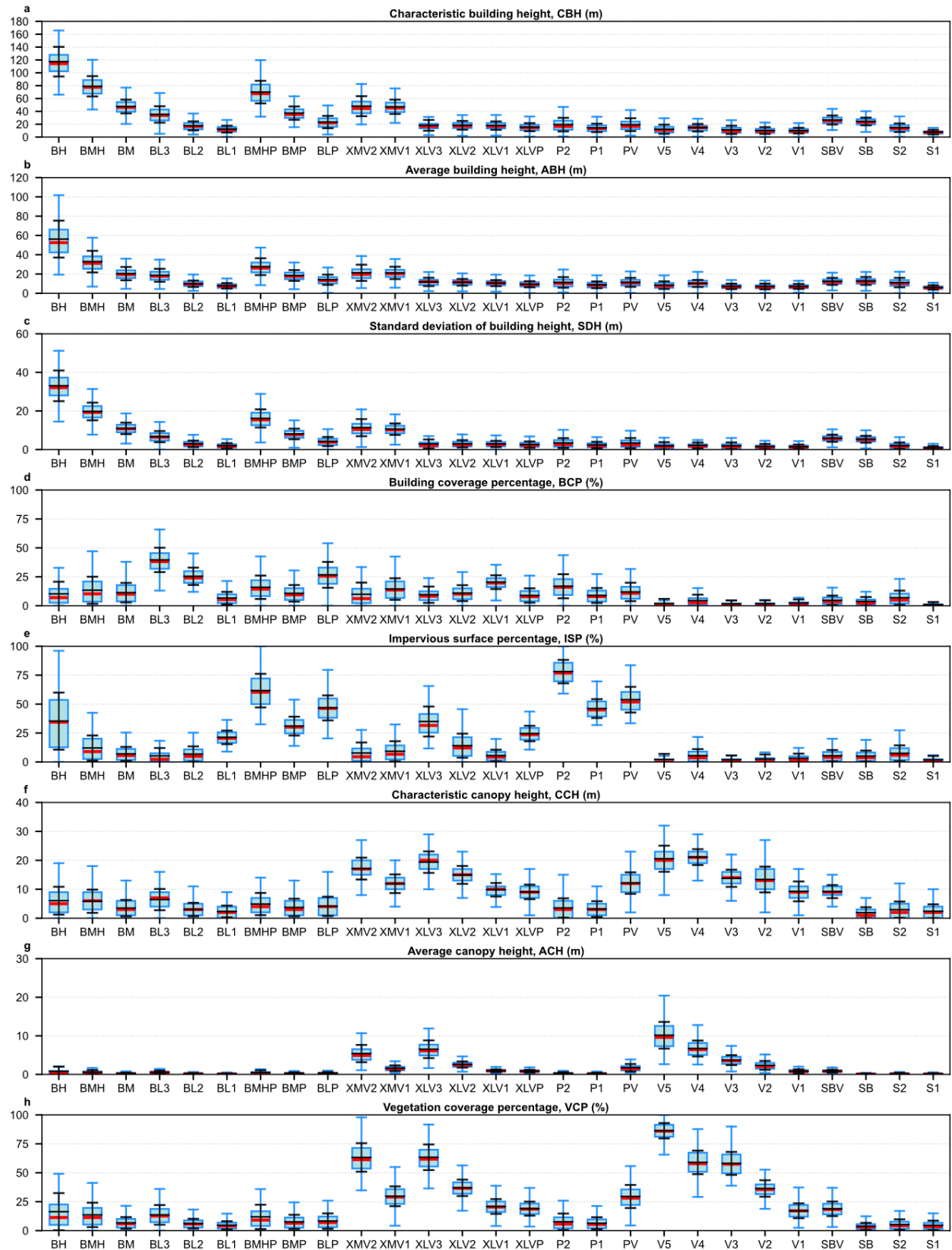

**Fig. S2: Normalized co-occurrence matrix of the DUEZ and LCZ classifications for global urban surfaces.**
The sum of each column is normalized to 100% to indicate the distribution of a specific LCZ type across various DUEZ types. This analysis includes 26,617,213 tiles in total at 500-m resolution.

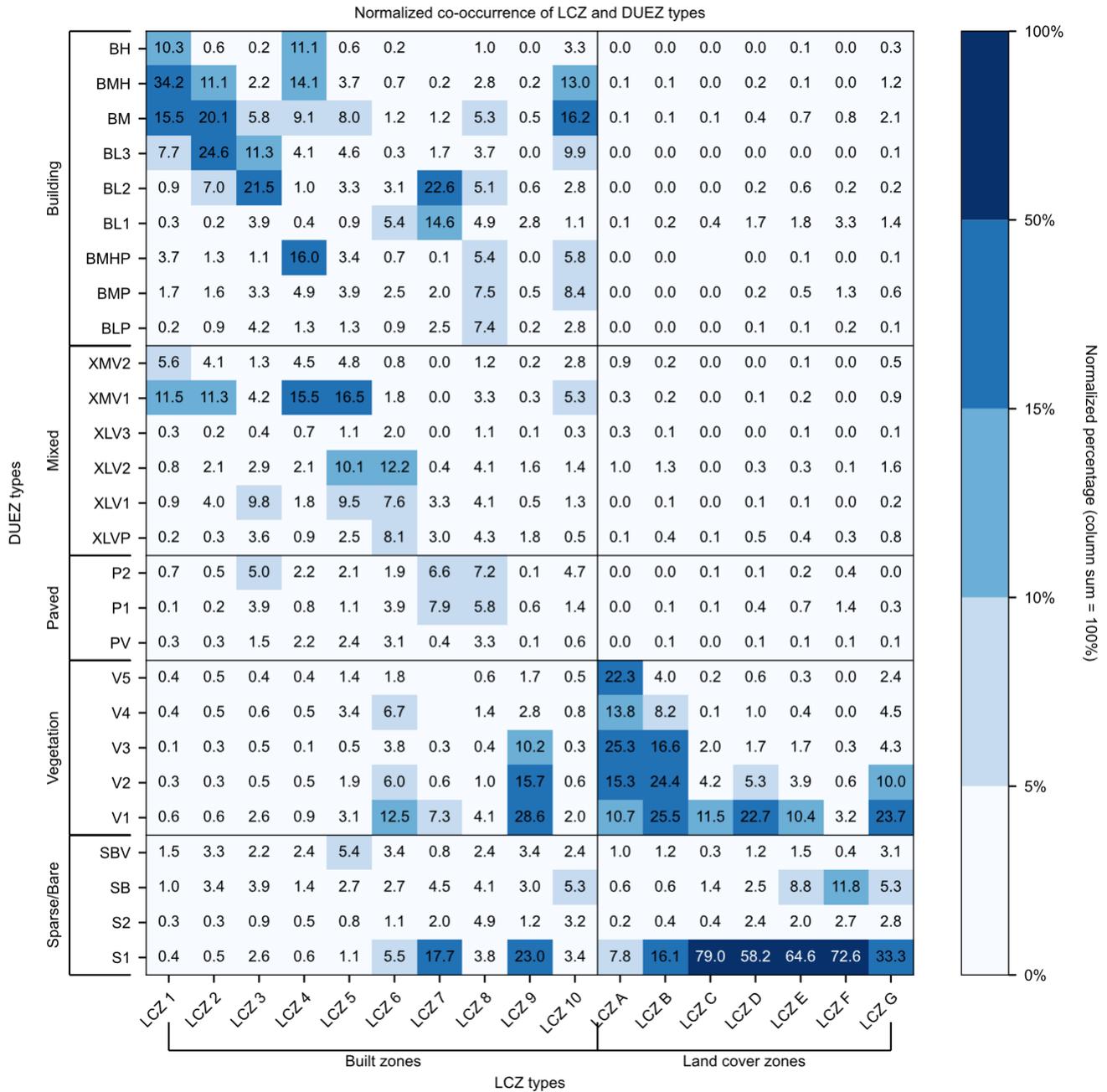

**Fig. S3: Global relative occurrences of individual DUEZ types in each of the nine urban textures.** Numbers indicate the standardized residuals that quantify how observed DUEZ-type counts deviate from counts expected if DUEZ types and urban textures were independent of each other; values exceeding 2 indicate a significant deviation from independence ($\chi^2$-test $p$-value < 0.001). Relative enhanced (red) and decreased (blue) presence of DUEZ are color-coded. The total number of DUEZ pixels analyzed is 1,992,332.

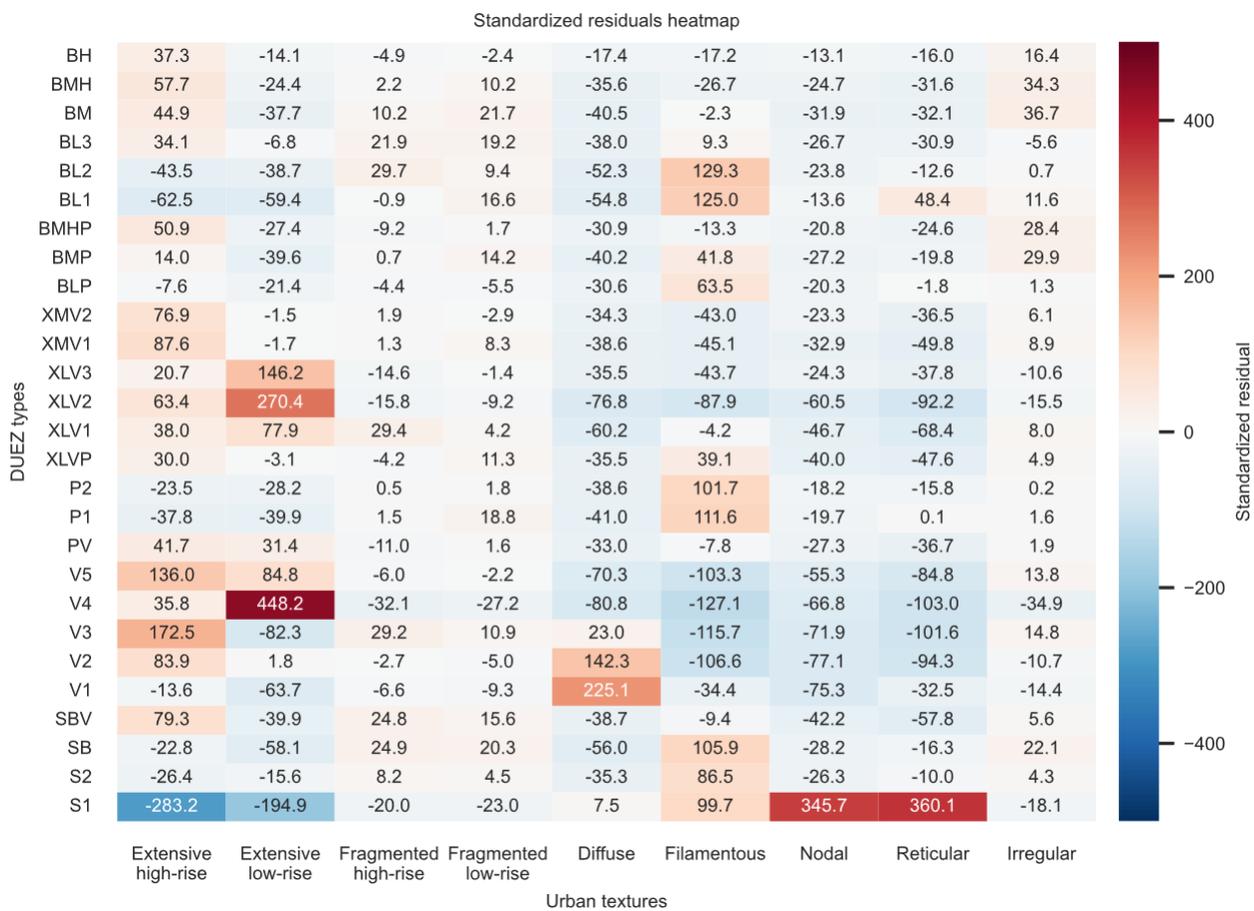

**Fig. S4: Spatial adjacency and assembly logic within distinct types of urban textures.**
**a**, extensive high-rise. **b**, extensive low-rise. **c**, fragmented high-rise. **d**, fragmented low-rise. **e**, diffuse. **f**, filamentous. **g**, nodal. **h**, reticular. **i**, irregular. Numbers in cells represent the percent probability of finding a specific DUEZ type (vertical axis) adjacent to a central DUEZ type (horizontal axis).

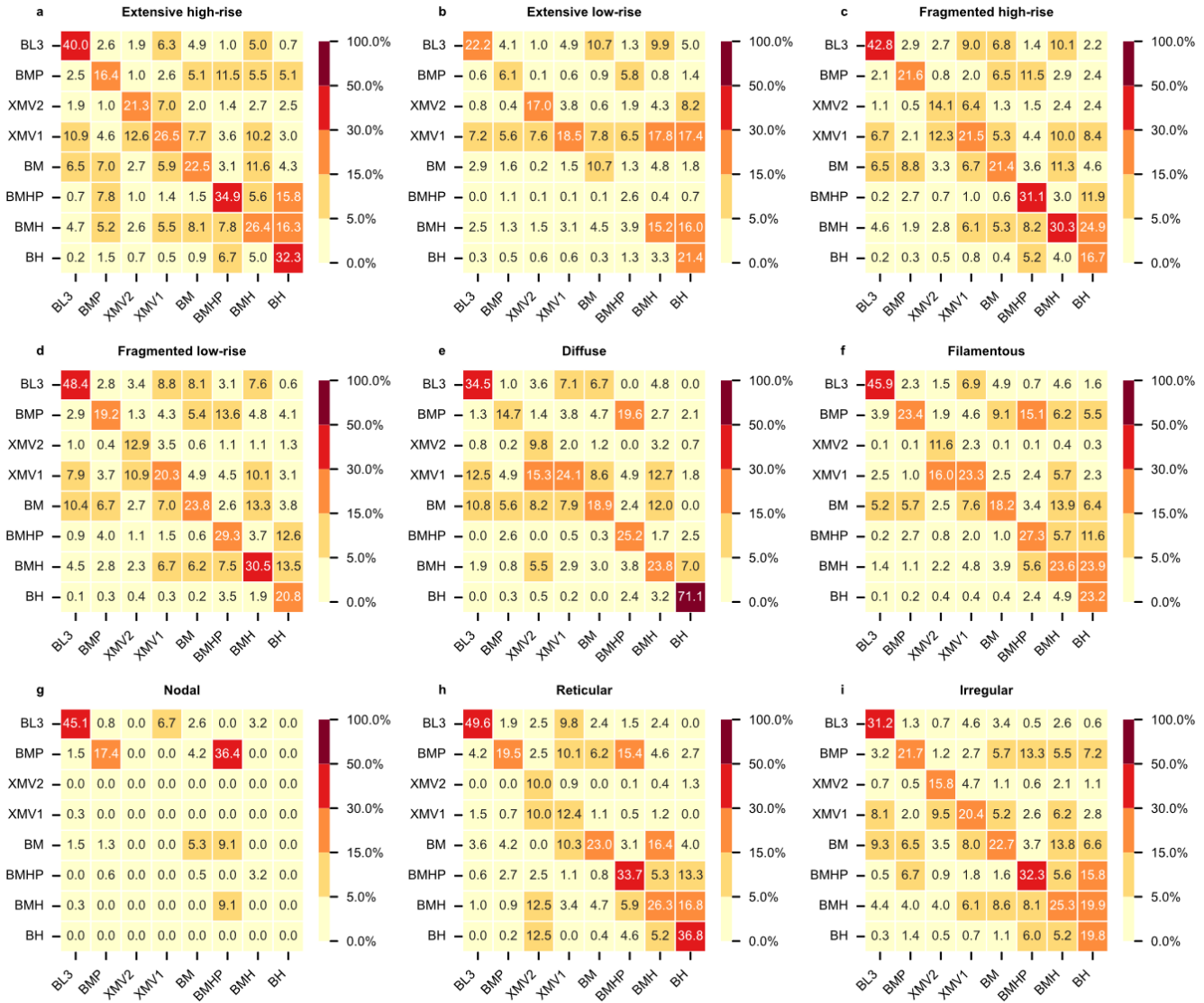

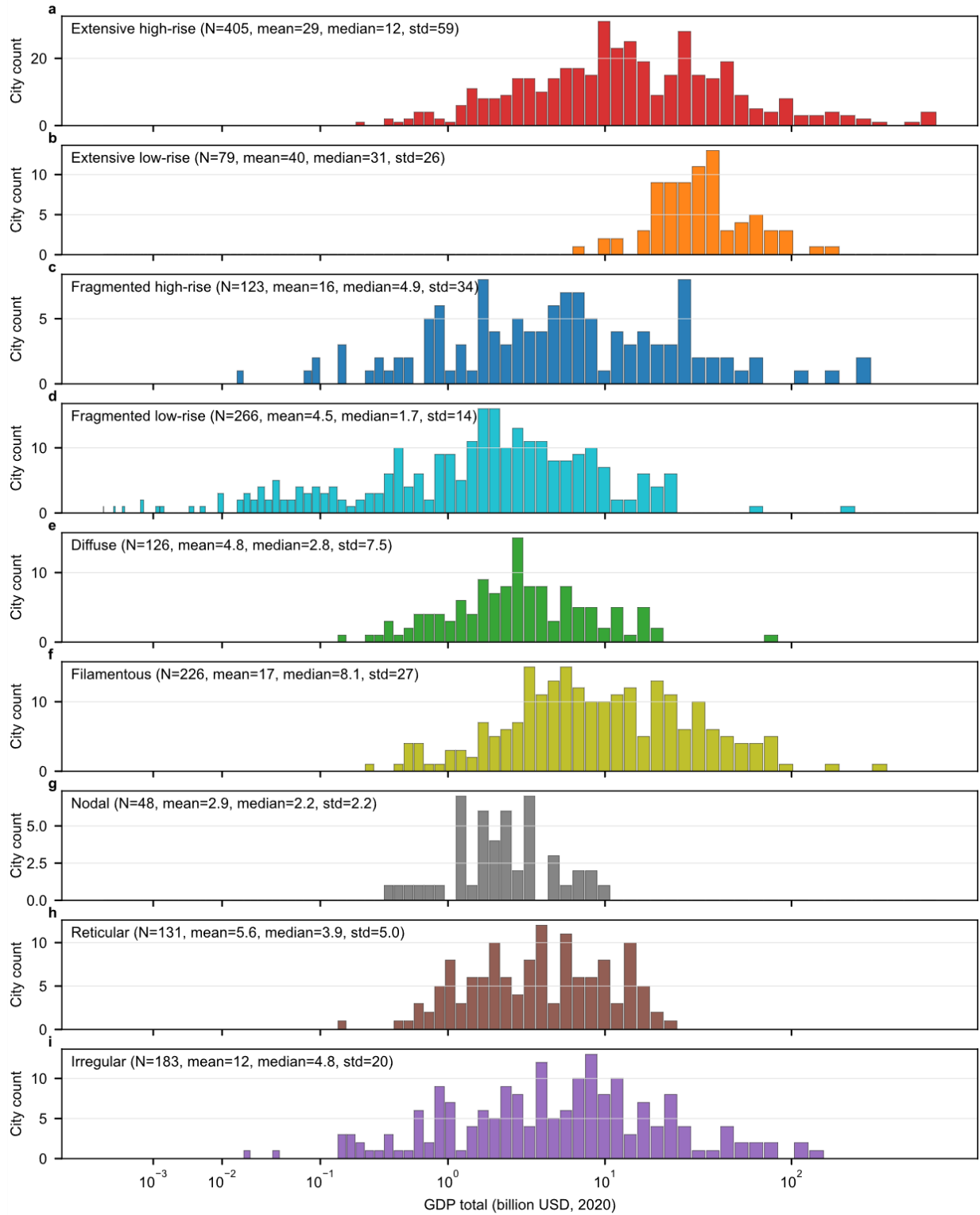

Fig. S5: Histogram of Gross Domestic Product (GDP, in billion USD for the year 2020) in 1,600 global cities classified by their urban textures. **a**, extensive high-rise. **b**, extensive low-rise. **c**, fragmented high-rise. **d**, fragmented low-rise. **e**, diffuse. **f**, filamentous. **g**, nodal. **h**, reticular. **i**, irregular. Each city is defined here as a 25 km × 25 km area to exclude the confounding impacts of city size on the analysis. The number of cities (N) and the mean and median GDP values are shown inset.

**Fig. S6: *P*-values from Dunn's *post-hoc* test for the GDP differences of 1,600 global cities across different urban textures.**
Green cells indicate that the GDP differences between the two urban textures are statistically significant.

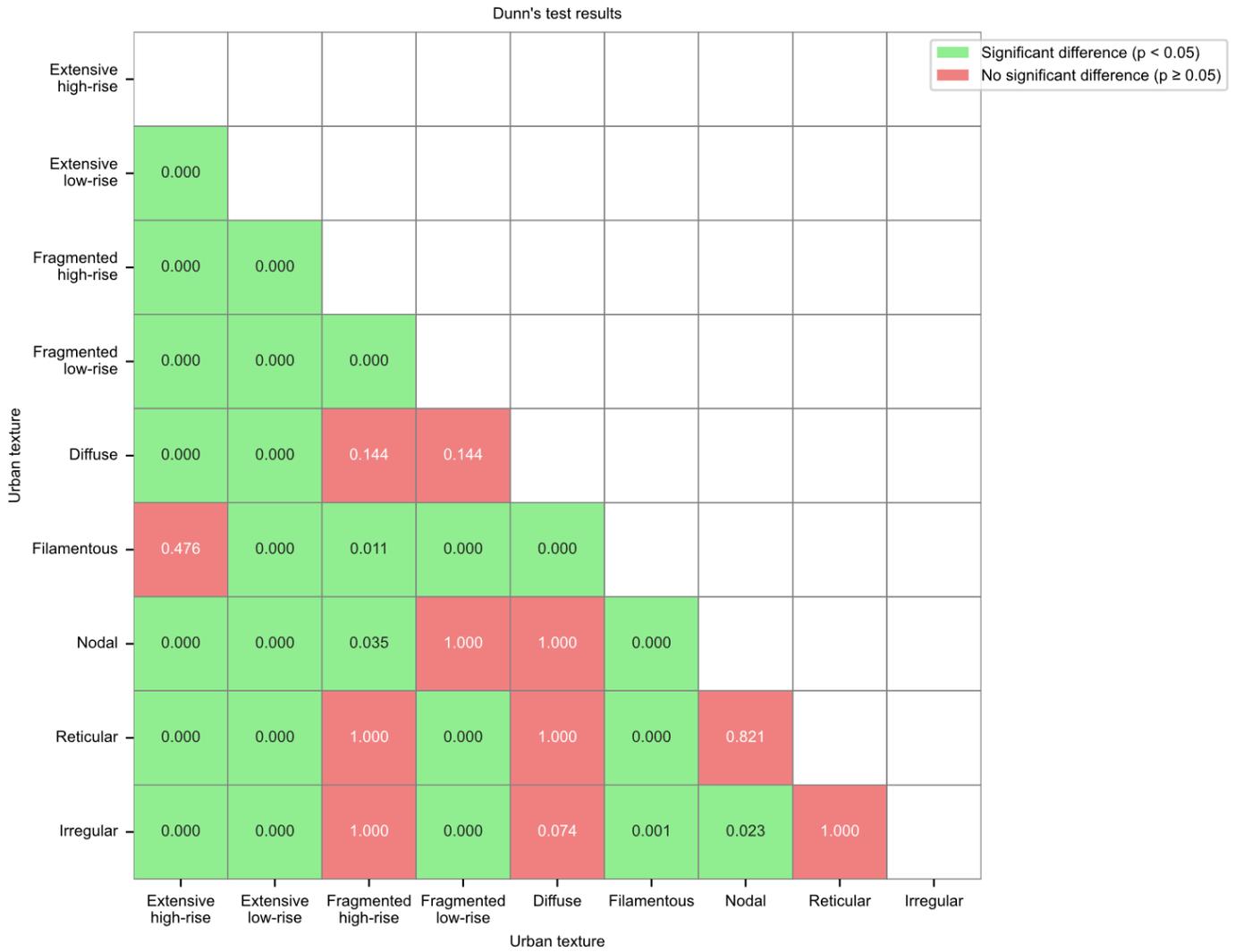

**Fig. S7: Pearson correlation of determination ($R^2$) among the eight urban surface parameters in the training dataset used for unsupervised DUEZ clustering.** Colored grid cells indicate statistically significant $R^2$ (two-tail $t$-test, $p$-values < 0.05).

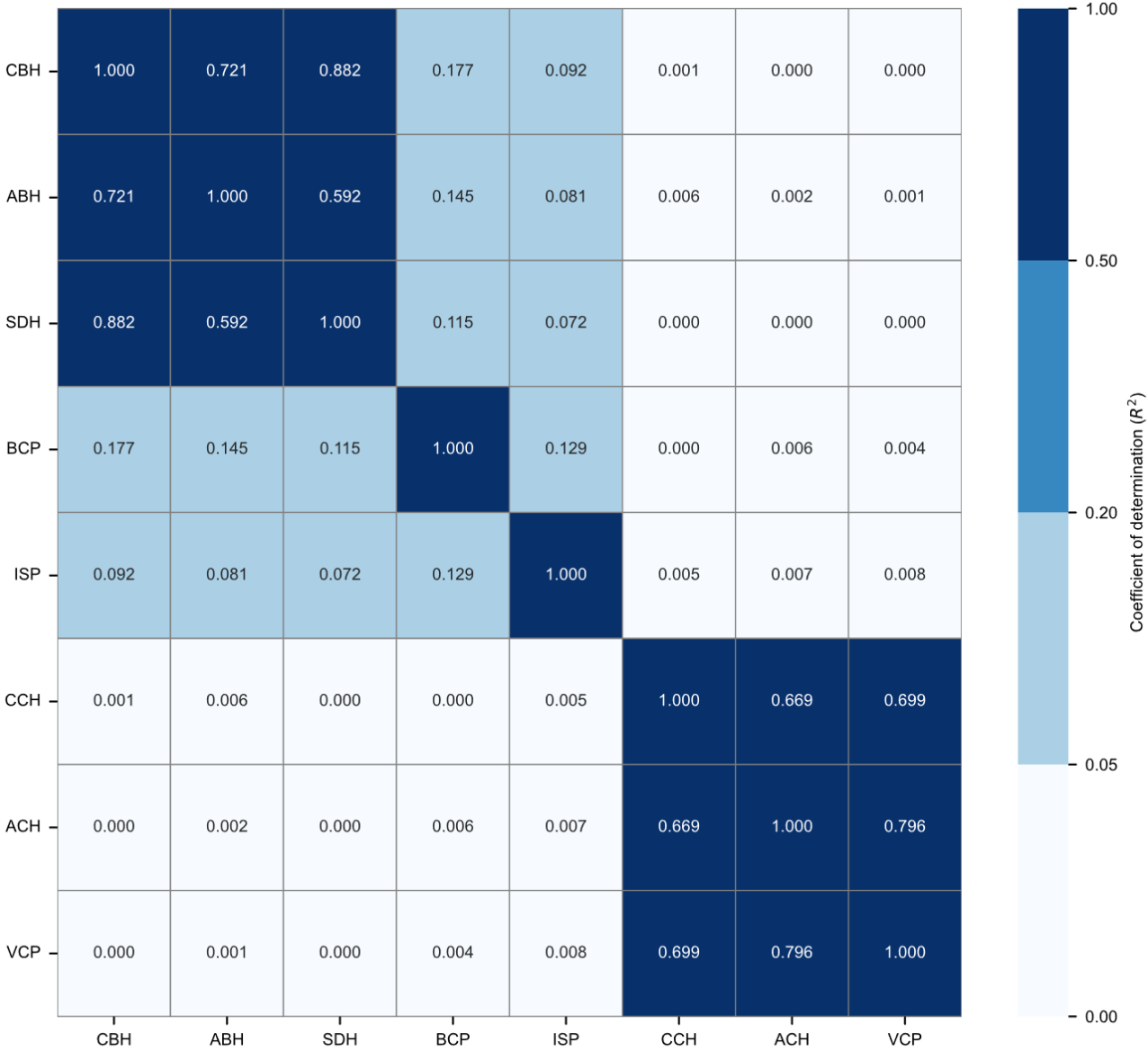

**Table S1. The eight surface parameters used in the unsupervised clustering of DUEZ.**

| Parameter | Description | Definition | Weight in Clustering |
|---|---|---|---|
| **CBH** | Characteristic building height (m) | The 98$^{th}$ percentile of the building heights weighted by the building area in the tile | 2 |
| **ABH** | Average building height (m) | The mean value of the building heights weighted by the building area in the tile | 2 |
| **SDH** | Standard deviation of building height (m) | The standard deviation of the building heights weighted by building area in the tile | 2 |
| **BCP** | Building coverage percentage (%) | The percentage of floor area covered by buildings in the tile | 1 |
| **ISP** | Impervious surface percentage (%) | The percentage of impervious surface area in the tile | 1 |
| **CCH** | Characteristic canopy height (m) | The 98$^{th}$ percentile of canopy height in the tile | 1 |
| **ACH** | Average canopy height (m) | The mean canopy height in the tile | 1 |
| **VCP** | Vegetation coverage percentage (%) | Percentage of area covered by canopy with height ≥ 1 m in the tile | 1 |